# AMBIENT BEAM MOTION AND ITS EXCITATION BY "GHOST LINES" IN THE TEVATRON*

V. Shiltsev, FNAL, Batavia, IL, USA

## Abstract

Transverse betatron motion of the Tevatron proton beam is measured and analyzed. It is shown that the motion is coherent and excited by external sources of unknown origins. Observations of the time-varying "ghost lines" in the betatron spectrum are reported.

## INTRODUCTION

Motion of the accelerator components, most notably, quadrupole magnets, results in the beam orbit movements and can lead to a significant deterioration of the collider performance in the Tevatron. The mechanism depends on the frequency. At very high frequencies comparable or higher than the betatron frequency $f_0 (1-\nu) \approx 19.7$ kHz fluctuations of the magnetic fields $\delta B(t)$, e.g. due to quadrupole magnet displacements $x(t)$, produce transverse kicks $\delta\theta(t) = \delta B(t)el/Pc = x(t)/F$, where $l$ is length of the element, $F$ is the focusing length, that leads to the beam emittance growth at the rate of [1]:

$$\frac{d\varepsilon_x}{dt} = \gamma \frac{f_0^2}{F^2} \sum_{k=1}^{N_q} \sum_{n=-\infty}^{\infty} \beta_k S_x(f_0(\nu-n)) \quad (1)$$

where $f_0$ is the revolution frequency, $\gamma$ is the relativistic factor, $\nu$ is the tune, $S_x(f)$ is the power spectral density of the quadrupole motion $x$, $N_q$ is a total number of quadrupole focusing magnets, and $\beta_k$ is the beta function at the location of the element. At much lower frequencies, $f \ll f_0$, the kicks lead to distortion of the closed orbit of the beams:

$$X_{COD}(s) = \frac{\sqrt{\beta(s)}}{2\sin(\pi\nu)} \sum_{k=1}^{N_q} \sqrt{\beta_k(s)} \theta_k \cos(\varphi(s) - \varphi_k + \pi) \quad (2)$$

where $s$ is location along the ring, and $\varphi(s), \varphi_k$ are betatron phases at the locations of the observation point and at the source of the $k$-th magnet. At very low frequencies, hours to years, the quadrupole magnet displacements are often governed by the *"ATL law"* [2,3] according to which the mean square of relative displacement $dX^2$ of the points separated by distance $L$ grows with the time interval between measurements $T$ as:

$$< dY^2 > = A\,TL \quad (3)$$

where $A$ is a site dependent constant of the order of $10^{-5\pm1}$ $\mu m^2/(s \cdot m)$, and brackets $<...>$ indicate averaging over many points of observations distanced by $L$ and over all time intervals equal to $T$. Such a wandering of the accelerator elements takes place in all directions. Corresponding average

* Work supported by the U.S. Department of Energy under contract No. DE-AC02-07CH11359

closed orbit distortion over the ring with circumference $C$ is equal to [4]:

$$< X_{COD}^2(s) >= \frac{\beta(\beta_F + \beta_D)}{8F^2 \sin^2(\pi\nu)} ATC = \kappa ATC \quad (4)$$

where FODO lattice structure is assumed, $\beta_F, \beta_D$ are beta functions at the focusing and defocusing lenses, and numerical coefficients $\kappa \approx 3$ and $A = (2-5) \times 10^{-6}$ $\mu m^2/(s \cdot m)$ [5] for the Tevatron.

The Tevatron orbit drifts from empirically found "good" orbits, significant changes in the betatron frequencies may occur that lead to (usually) higher losses of antiprotons and protons. At the injection energy of 150 GeV when the beams are several mm wide, orbit motion of about a mm leads to losses of the beams at several known places with tight aperture. At the energy of experiment, 980 GeV/beam, beam position in the RF cavities affects stability of high-intensity proton beam, e.g. the power of coherent beam oscillations goes up if the beam is too far off center. Also, oscillations of the RF cavities at synchrotron frequency (85 Hz at 150 GeV and 35 Hz at 980 GeV) are of concern for driving longitudinal emittance growth due to microphonic effects [6]. Large scale long-term drifts of the orbit can be corrected by dipole correctors and regular realignment of the magnets – usually during annual shutdown periods – helps to keep the corrector currents under the limit of 50 A.

## BETARON MOTION AMPLITUDE

Several instruments are being used to detect orbit motion in the Tevatron.

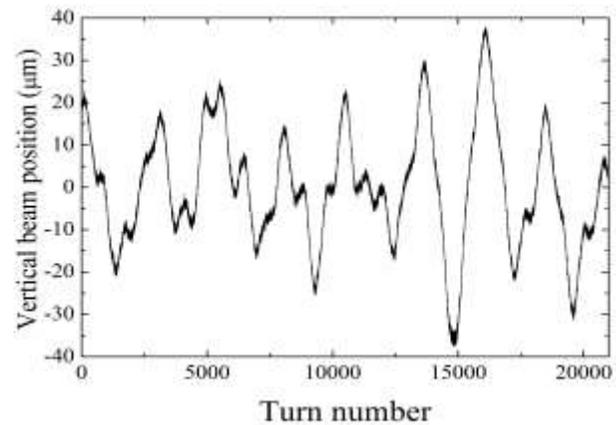

Fig.1: Vertical position of the proton bunch #11 at the beginning of HEP store #6214 (October 2008), measured at the VB11 location with $\beta_y$=900 m.

The most challenging is direct measurement of the minuscule betatron oscillations. Several special instruments were built for the purpose of detecting natural (ambient) beam oscillations and, therefore, determination of the betatron frequencies without additional excitations.

Various techniques were employed, including 3D-BBQ (Direct Diode Detector Baseband tune measurement system (3D-BBQ) [7] and the Digital Tune Monitor (DTM) which uses 16 bits 100 MHz ADC's for measuring the tunes on a bunch-by-bunch basis [8]. Very high precision system employing a fast digital scope (Agilent Acqiris, 10bit, 8GS/s) for measurements of the turn by turn vertical centroid positions of individual bunches has been devised and used for digitizing signals from the plates of the VB11 BPM in the large vertical beta function location that translates into better S/N ratio [9]. The system employs variable attenuators for compensating the beam position offset and phase shifters synchronized within some 10 ps to minimize common mode. As the results, subtraction of the two signals by an RF hybrid provides about 44 dB of common mode suppression. Fig.1 shows 21,400 turn (0.44 s) record of the vertical beam position at the location.

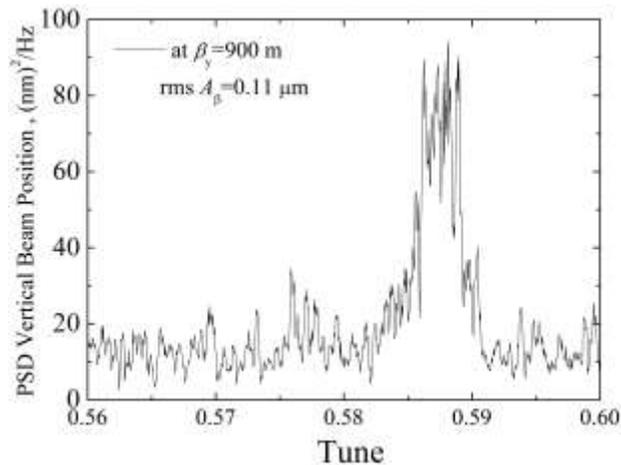

Fig.2: Power spectral density of the vertical betatron oscillations (FFT of the data presented in Fig.1).

The FFT of the data reveal significant excess of the signal at the betatron tune line over the noise as shown in Fig.2. To calculate properly the rms amplitude of the betatron oscillations one has to a) filter all harmonics except at the tune range 0.4-0.5 (FFT filter); b) make FFT of the remaining signal; c) determine noise level (pedestal); d) subtract it from the signal at the betatron line; f) determine the signal level. At the end, the rms amplitude of the betatron oscillations is found to about 110 nm. Note that the amplitude significantly varies in stores and from store to store (see next chapter) and often is 3-8 times smaller, that yields some 5-25 nm range of typical rms betatron motion amplitudes at the average beta function location with $\beta_y \approx 50$ m. Spectrum of the vertical orbit motion at frequencies 2-1000 Hz is shown in

Fig. 3. It scales approximately as $\propto 1/f^3$ and is dominated by the low frequency beam motion. The strongest lines are the harmonics of 60 Hz main power. The 15 Hz and the 0.45 Hz components can be explained by the effects of the fast cycling Booster synchrotron and the Main Injector on the power distribution systems at FNAL.

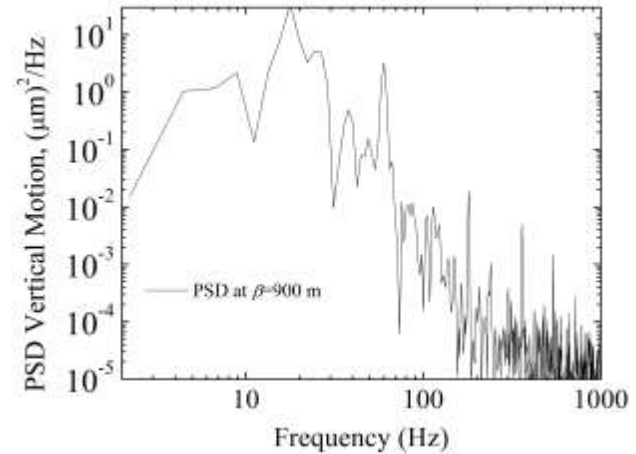

Fig.3: Low frequency power spectral density of the vertical orbit oscillations.

## "GHOST LINES"

As noted above, the beam oscillation spectrum varies a lot, depending on conditions. The fastest and most widely operationally used instrument to monitor them is 21MHz Schottky detector [10]. The 21.4 MHz Schottky system is used to measure the horizontal and vertical tunes of the proton beam (antiproton signal is attenuated by 20 dB) without the possibility of gating on individual bunches. It has high resolution of about 0.0001.

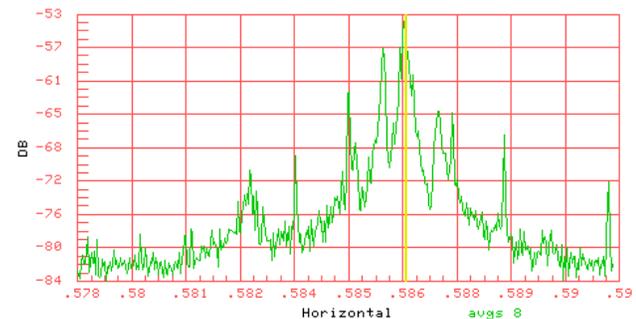

Fig.4: Horizontal 21 MHz Schottky spectrum of proton beam measured at 980 GeV measured 09/29/2002.

A typical Schottky spectrum from the 21.4 MHz pickups measured during collisions is shown in Fig.4. The spectrum contains many peaks, a set of lines around 0.586 are horizontal betatron tune lines (the vertical tune line was at 0.574 and is off the scale of the plot). Other remarkable peaks at 0.582, 0.584, 0.589 and 0.591 are what is called "ghost lines" corresponding to beam excitations of unknown origins. These lines were present in the Schottky spectra since early days of the Collider Run II and their most puzzling feature is that they are not stable but move around the frequency spectrum without any obvious reason. When they propagate through the vertical or horizontal oscillation frequencies, the power in the betatron motion goes up significantly. Fig.5 shows some 1.5 hour record of the Schottky spectrum during one of the colliding stores. Brighter lines are 0.596 and 0.585 indicate betatron bands, while several ghost lines were crossing them – coming both from below and from above. Note that their frequencies not only drift but also wiggle by as much as 0.005-0.02 with periods of about 15-20 minutes.

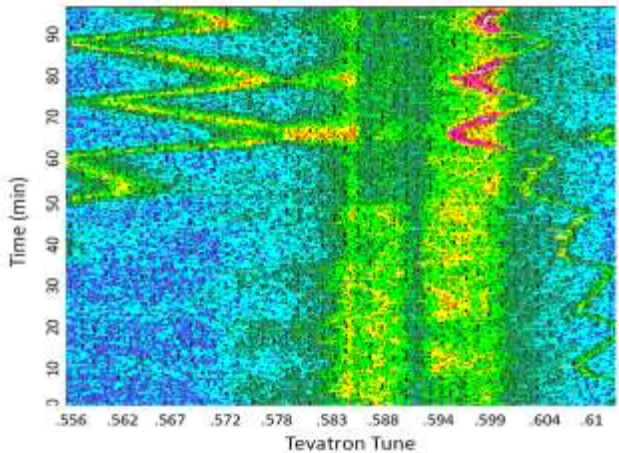

Fig.4: Horizontal 21 MHz Schottky spectrum of proton beam measured at 980 GeV (see text).

Fig.6 shows variation of the total Schottky power in the frequency band covering some 0.6 units of the tune space during a 14 hours long colliding store. One can see that the horizontal power varies by 18 dB , equivalent to the rms amplitude variation by a factor of 8.

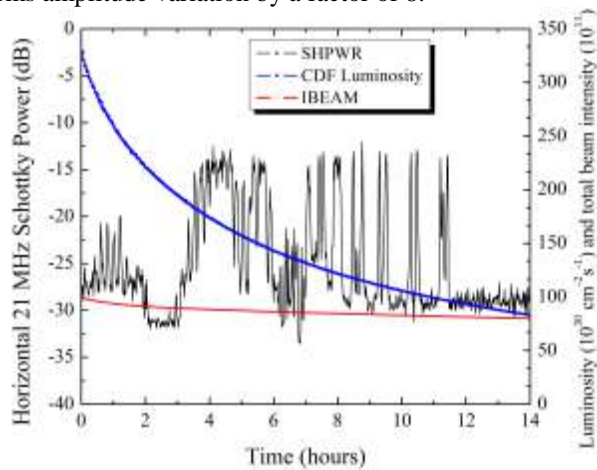

Fig.6: Variation of the horizontal 21 MHz Schottky power during 14-hours colliding run #8548 (March 9, 2011).

## DISCUSSION AND SUMMARY

The direct measurement of the rms betatron oscillations amplitude estimates it at about 110 nm at $\beta_y \approx 900$ m. Correspondingly, at the amplitudes at the average beta function location with $\beta_s \approx 50$ m is about 25 nm. Given that such direct measurements with clearly observable betatron peak were not repaatedly reproducible, one can conclude that well know "ghost lines" are the reason for that – as they are come and go without any obvious regularity. Our analysis of these "ghost lines" shows that a) besides slow motion across frequencies, they also exhibit oscillatory movements with period varying from 15-20 min to few hours; b) for the stores analysed, the lines add about factor of 2 to average – over colliding store duration- Schottky power in the betatron bands. The latter allows to estimate that they contribute about half to the previously determined the rms normalized emittance growth rate of some 0.06 $\pi$ mm mrad/hr [11].

The Tevatron "ghost lines" look very similar to infamous "humps" recently observed in the LHC. Those "humps" are unwanted oscillations seen repeatedly in the LHC beams (mostly in the vertical plane) and also believed to be caused by external excitations [12].

I would like to thank A.Valishev, G.Stancari, D.Still and J.Annala for providing raw BPM data and for many useful discussions on the subject of the "ghost lines".